\documentclass[aps,twocolumn,groupedaddress,superscriptaddress,floatfix,pra]{revtex4-2}
\usepackage{amsmath}
\usepackage{mathtools}
\usepackage{amssymb}
\usepackage{graphicx}
\usepackage{textcomp}
\usepackage{bm}	
\usepackage{soul}
\usepackage{caption}
\usepackage{subcaption}

\usepackage[usenames,dvipsnames]{color}

\usepackage[braket, qm]{qcircuit}

\begin{document}

\title{Cooperative isentropic charging of hybrid quantum batteries.}

\author{Yohan Vianna de Almeida}
\affiliation{Instituto de F\'isica, Universidade Federal do Rio de Janeiro, CP68528, Rio de Janeiro, Rio de Janeiro 21941-972, Brazil}

\author{Tiago F. F. Santos}
\affiliation{Instituto de F\'isica, Universidade Federal do Rio de Janeiro, CP68528, Rio de Janeiro, Rio de Janeiro 21941-972, Brazil}

\author{Marcelo F. Santos}\email{Corresponding author: mfsantos@if.ufrj.br}
\affiliation{Instituto de F\'isica, Universidade Federal do Rio de Janeiro, CP68528, Rio de Janeiro, Rio de Janeiro 21941-972, Brazil}

\pacs{xxxx, xxxx, xxxx}                                         
\date{\today}
\begin{abstract}
Quantum batteries are quantum systems used to store energy to be later extracted by an external agent in the form of work to perform some task. Here we study the charging of a hybrid quantum battery via a collisional model mediated by an anti-Jaynes Cummings interaction obtained from an off-resonant Raman configuration. The battery is made of two distinct components: a stationary infinite dimensional single quantum system (e.g. an harmonic oscillator) and a stream of small dimensional ones (e.g. qutrits). The charging protocol consists of sequentially interacting the harmonic oscillator with each element of the stream, one at a time, under the action of an external energy source and the goal is to analyze how the charging of both the harmonic oscillator and the qutrits is affected by the correlation properties of the stream.
\end{abstract}

\maketitle

\section{Introduction}
With the development of new quantum technologies, it is a new fundamental task to understand the mechanisms of energy exchange in the quantum realm, and the role played by quantum resources in microscopic devices. In this framework, a topic that has attracted lots of attention in recent years is the charging of quantum batteries. Researchers on this topic explore how quantum effects, such as quantum coherence and entanglement, can be useful to speed up the charging and increase the input power and the work extraction in quantum batteries~\cite{karen2013, andolina2018, Farina2019, andolina2019, carrega2020, caravelli2020, sergi2020, liu2021, stella2021, Mondal2022, Gyhm2022, Ghosh2022, Kang2022, raf2023, dou2023}. This field has also met experimental advancements in the production of such microscopic devices~\cite{ Horne2020, Quach2022, Chang2022, joshi2022}. Quantum batteries can be considered standing still systems from which one can draw stored energy whenever needed, or they can be considered moving active media used to feed energy into other systems. 

Quantum optics is one of the research fields where the exchange of energy between external sources, heat bathes, and microscopic quantum systems has been extensively studied for the last 60 years~\cite{ScovilDuBois, 60years, 60years2, 60years3, 60years4, 60years5}. Therefore, it comes as no surprise that it has also been a powerful testbed for the concepts and the development of the recent research area of quantum thermodynamics. In fact, the maser, a system that can be considered as the experimental origin of quantum optics, is also one of the first quantum heat engines analyzed as so~\cite{ScovilDuBois}. More recently, quantum optical setups have been used to understand or simulate the charging of quantum batteries and the implementation of microscopic heat engines ~\cite{SRQHE, AC, AC2, molmer, QD, Bera, gemme, mm, lossy}.

One of the advantages of exploring quantum optical setups in the study of quantum thermodynamics is the possibility to design well controlled scenarios of distinct interacting systems where the roles played by each part is clearly defined and quantities such as work and exchanged heat can be quantified and, sometimes, even measured or at least simulated~\cite{sim, sim2, sim3, sim4}. Moreover, some of the most canonical setups involve the coupling of systems of different natures such as few atomic levels, addressed as d-level atoms, where d is typically two or three, coupled with one or more quantum harmonic oscillators such as modes of the electromagnetic field or vibrations of those same atoms in some trapping potential. The control level over these interactions allow for a multitude of applications including the interplay of different roles played by each system. For example, two-level atoms sequentially interacting with a cavity field can play the role of a heat reservoir for the field but one that can be eventually monitored and from which information can be extracted, allowing, for example, for the elegant design of Maxwell Deamon's setups where the Deamon can be actually probed~\cite{Daemon, Daemon2, Daemon3, Daemon4}. In this scenario, the field may be considered the quantum battery where energy is stored. On a reserved role, leaking cavity modes can play the role of effective high or low temperature reservoirs for trapped atoms (or ions) but, once again, reservoirs that can be monitored and from which information can be extracted, leading to the eventual preservation of coherence or entanglement in the atomic systems, the testing of fluctuations theorems or even the implementation of universal and efficient quantum computation~\cite{qcomput, qcomput2, qcomput3, qcomput3, qcomput4, qcomput5}. Now, the trapped atom plays the role of a quantum battery. Usually, as it is the case in all these examples, one of the interacting systems is the energy store (or battery) and the other is the energy provider or extractor~\cite{liu2021, stella2021, charger, charger2, charger3, charger4}. 

\begin{figure}[h!]
    \centering
    \includegraphics[scale=0.5]{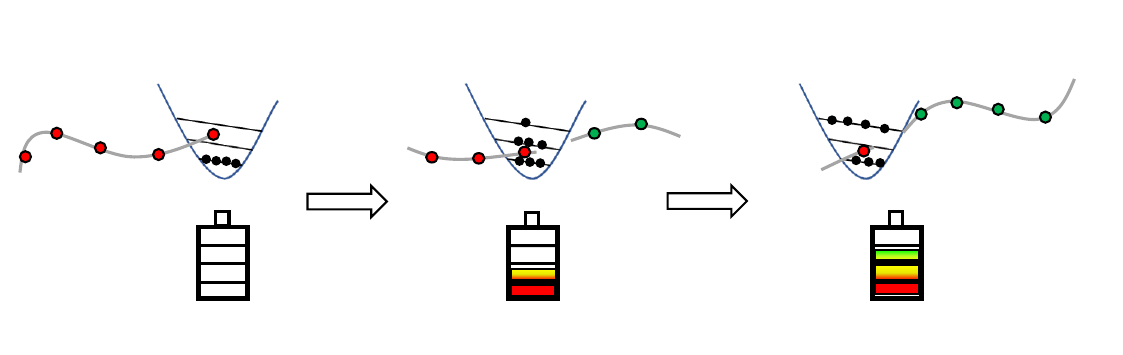}
    \caption{As the stream of atoms pass through the single-mode electromagnetic field, Rabi oscillations are induced between different doublets, thus charging both the chain and the field altogether.}
    \label{stream}
\end{figure}

In this work, we will employ standard quantum optical techniques to focus on a different approach: we will consider the entire interacting system as a hybrid quantum battery, one that has a standing still component of infinite dimension (e.g. a quantized mode) and an assembling line of effective two-level batteries built upon a stream of 3-level atoms, as depicted in Figure (\ref{stream}). We will study their concomitant charging, particularly focusing on the effects of quantum or classical correlations previously prepared into the assembling line. More specifically, each atom of the stream interacts with the field for a finite time interval during which they are both coherently pumped by, and extract work from, an external source. In each step, the total drained energy is split into the atom and the quantized mode, and the interaction is reset for the following atom. We focus our study on the efficiency of this work extraction not only in terms of the atom-field coupling parameters but, more important, in terms of the correlation properties of the assembling line. Our main goal is to understand how quantum or classical correlations previously created in the atoms affect the charging process of each system, which is an ongoing debate in the modern literature~\cite{correl1, correl2, correl3}.

It is worth noticing that, from the thermodynamics point of view, we will concentrate our results on an isentropic exchange of energy, where the external source is coherently coupled to both systems and those are isolated from external heat reservoirs so that the overall time evolution is unitary. 
The paper is organized as follows: in section II we briefly introduce the dynamics of the interaction of each atom of the stream and the quantized mode. Then, in section III, we pass to the description of the charging of both the atomic chain and the quantum field batteries and show how the energetic gain peaks together for both systems. This result is simple enough to be derived analytically. In section IV, we discuss the roles that correlation and entanglement play in the protocol of charging the quantum field. We analyze each kind of evolution for their energy gain and stored ergotropy~\cite{Erg}, and compare the relative performances with the resource consumption for creating each initial state. Finally, we present the conclusions and interesting limits for the operation of the protocol.

\section{System's description}
One quantum optical interaction that allows for the simultaneous energy transferring to both finite and infinite systems is the so-called anti-Jaynes-Cummings Hamiltonian where a two-level atom and a quantized harmonic oscillator can both absorb or emit one quantum of energy at a time. Because this Hamiltonian does not preserve the overall number of excitations of the system, it typically does not occur spontaneously and it has to be engineered in most situations. Here, we will use the same setup presented in \cite{MKR, TYM} where the anti-Jaynes-Cummings interaction derives from an off-resonant Raman configuration.

The system under study is composed of a single mode of a quantized electromagnetic field of frequency $\omega_q$, a chain of $K$ three-level atoms, and a classical power source (PS) which generates an oscillatory potential of frequency $\omega_L$. The external PS pumps energy into the system by intermediating the coupling between each atom $k$ and the quantized mode, which takes place for a finite duration time $\Delta\tau_{k}$. Each three-level atom is described by its free energy eigenstates \{$\ket{g}, \ket{e}, \ket{m}$\}, with $E_m>E_e>E_g$, and $E_j = \hbar \omega_{j}$, and the atomic levels are coupled to the classical PS and the quantized mode according to $H = H_0+H_I$, where $H_0=\sum_j E_j|j\rangle\langle j|+\hbar \omega_q b^\dagger b$ and
\begin{equation}\label{CH}
H_I = \Omega_L (\sigma_{gm}e^{i\omega_L t}+\sigma_{mg}e^{-i\omega_L t}) +g_q (\sigma_{em}b^\dagger + \sigma_{me}b). 
\end{equation}
Here, $\sigma_{mn} = |m\rangle \langle n|$ and the couplings to both the classical PS and the quantum field are off-resonant by a large gap $\Delta$, meaning $\Delta = \omega_{mg} - \omega_{L} = \omega_{me} - \omega_{q} >> \Omega_{L}, g_{q}$ ($\omega_{ij} = \omega_{i}-\omega_{j}$). Under these conditions, it has been shown in~\cite{MKR, MFS, Nicim} that we can adiabatically eliminate level $\ket{m}$ from the dynamics on each atom, creating an effective interaction between levels $\ket{g}$ and $\ket{e}$, given by the Hamiltonian:
\begin{equation}
    H_{eff} = -\dfrac{g_{q}^{2}N}{\Delta}\sigma_{gg} - \dfrac{g_{q}^{2}\hat{b}^{\dagger}\hat{b}}{\Delta}\sigma_{ee} + \dfrac{\Omega_{L}g_{q}}{\Delta}(\sigma_{ge}\hat{b}+\sigma_{eg}\hat{b}^{\dagger}).
    \label{Heff}
\end{equation}
The first two terms of Eq. (\ref{Heff}) correspond to d.c. Stark shifts to level $|g\rangle$ ($|e\rangle$) due to the dispersive coupling to the adiabatically eliminated level $|m\rangle$ through the P.S. (quantized field). The third term, of the same order of the first two, responds for an effective interaction between levels $|g\rangle$ and $|e\rangle$ mediated by the P.S., the quantized field and, once again, level $|m\rangle$. Notice that, following the derivations in~\cite{MKR, MFS}, $H_{eff}$ also includes a small correction $\hbar\Delta_{eg}^{N} = \hbar\dfrac{\Omega_{L}^{2}-g_{q}^{2}N}{\Delta}$ to the energy difference between levels $\ket{g}$ and $\ket{e}$, where $N$ is a tunable parameter. This correction can be physically accounted by a small d.c. Stark shift to each atom which does not affect the conditions for the elimination of level $\ket{m}$.

As previously discussed, this effective Hamiltonian splits the total Hilbert space of each atom and the quantized field into doublets \{$\ket{gn},\ket{en+1}$\} (where $n$ is the number of photons in the quantized mode of the field) and draws energy from the classical potential to induce Rabi oscillations in each of them. It corresponds to an anti-Jaynes-Cummings (anti-JC) configuration, where a photon extracted from the external Power Source is split into two excitations, one for the atom and one for the quantized field. In a recent paper~\cite{TYM}, we have shown that this anti-JC configuration plus the vacuum of the quantized mode combine to allow for a full population inversion of the atom in an isentropic dynamics, regardless of its initial temperature.

As the evolution of each atom has already been discussed~\cite{TYM}, we now focus on the process of charging the whole chain and the quantum field. We work with a collisional model of non-interacting atoms, where the individual state of each atom in the chain, before the interaction with the field, is thermal, and we compare three different scenarios for the global state of the atomic chain: a) it is uncorrelated; b) it presents classical correlation; and c) it is an entangled state. In every case, each atom of the chain evolves unitarily with the quantum field under the effective Hamiltonian (\ref{Heff}) for a time $\Delta\tau_{k}$, where $k$ represents the interaction with the $k^{th}$ atom of the chain.

\section{Hybrid quantum battery charging protocol}
We assume that each atom in the streamline and the single-mode electromagnetic field ($E.M.$) are initially in a Gibbs state
\begin{equation}\label{gs}
    \rho_j = \frac{e^{-H_j/k_BT}}{Z},
\end{equation}
where $Z = \operatorname{Tr} e^{-H_j/k_BT}$ ($ j = atom, field$). At each collision, a single atom of the assembling line interacts with the $E.M.$ field for a finite time interval $\Delta \tau_k = \tau_k - \tau_{k-1}$, and energy is injected in both systems by the external P.S. represented by $\Omega_L$ in Eq. \eqref{CH}. After each interaction, the energy gained by the interacting atom is given by
\begin{equation}\label{uat}
    \Delta U_k^{atom} = \hbar \omega_{eg} S(\tau_k)
\end{equation}
and the energy gained by the $E.M.$ field is
\begin{equation}\label{uf}
    \Delta U^{field}_k = \hbar \omega_q S(\tau_k),
\end{equation}
where $U_k^j = \operatorname{Tr} (\rho_k^j H_0^j)$ and
\begin{equation}\label{Sk}
\begin{split}
    S(\tau_{k}) \equiv \sum_{n=0}^{\infty} A_{n}&\left[p_{g}^{T}p_{n}(\tau_{k-1}) - p_{e}^{T}p_{n+1}(\tau_{k-1})\right]\\ &\times sin^{2}\left(\dfrac{\Omega_{n}\Delta\tau_{k}}{2}\right).
\end{split}
\end{equation}

Here $\Omega_{n} = \dfrac{g_{q}\Omega_{L}}{\Delta}\sqrt{r^{2}(n+1-N)^{2} + 4(n+1)}$ is the Rabi frequency of oscillation for the doublet $\{|g,n\rangle,|e,n+1\rangle\}$, with $r\equiv\dfrac{g_{q}}{\Omega_{L}}$ the ratio between the couplings, $p_n (\tau_k)$ is the field's population after the $k^{th}$ collision, and $p_{g}^{T}=e^{-\beta\hbar\omega_g}$, $p_{e}^{T}=e^{-\beta\hbar\omega_e}$ are the atomic's initial population which are thermal by construction (each atom is individually in a thermal state before interacting with the field). Depending on the value of $r$, the selectivity amplitude factor $A_{n} \equiv \dfrac{1}{1+\frac{r^{2}(n+1-N)^{2}}{4(n+1)}}$ allows one to choose which doublet will influence more the dynamics by choosing a specific value $N$ for the d.c. Stark shift. In the limit of $r\gg 1$, choosing $N=n+1$ selects a single doublet to be involved in the energy exchange ($A_n$ goes to zero for every $n$ except for $n=N-1$)~\cite{MKR,MFS}. In the opposite limit of $r \ll 1$ the dynamics approaches that of a standard anti-J.C. interaction. In this case, all the doublets take part in the unitary evolution and the exact value of $N$ becomes irrelevant for the dynamics since both $A_n$ and the Rabi frequency $\Omega_n$ in each doublet become independent of $N$. Both regimes have been explored in the context of understanding quantum effects in the charging of quantum batteries in~\cite{TYM,MFSFB} but never related to any kind of correlation in the systems.

In our protocol, we operate with $r \ll 1$, which is the regime with the highest energy gain for single shot interactions, as shown in \cite{TYM}, and we choose the time interval $\Delta \tau_k$ in such a way that the energy gained after each interaction is maximized, i.e., the function $S(\tau_k)$, given by Eq. \eqref{Sk}, reaches its maximum value in each collision. Note that, by comparing Eqs. \eqref{uat} and \eqref{uf} it can be seen that the ratio between $\omega_{eg}$ and $\omega_q$ defines which system will have the highest energy gain after the collision. Also note that, in general, $\Delta \tau_k$ changes as the E.M. mode charges and higher $n$ doublets become more relevant. Finally note that, since charging only occurs during the interaction with each individual atom, maximizing the total charging of the batteries or the power efficiency of the protocol is equivalent.

To give an example of our charging protocol, we will consider an assembling line of atoms with K atoms prepared in a completely uncorrelated state
\begin{equation}
    \rho_{atom}^{total}(\tau_0) = \rho_{atom}^{(1)} \otimes \rho_{atom}^{(2)} \otimes ... \otimes \rho_{atom}^{(K)}.
\end{equation}
The figure of merit for this scenario is the energetic gain per collision, which we numerically calculate as a function of the number of atoms present in the colliding streamline.

\begin{figure}[h!]
  \centering
   \includegraphics[width=\columnwidth]{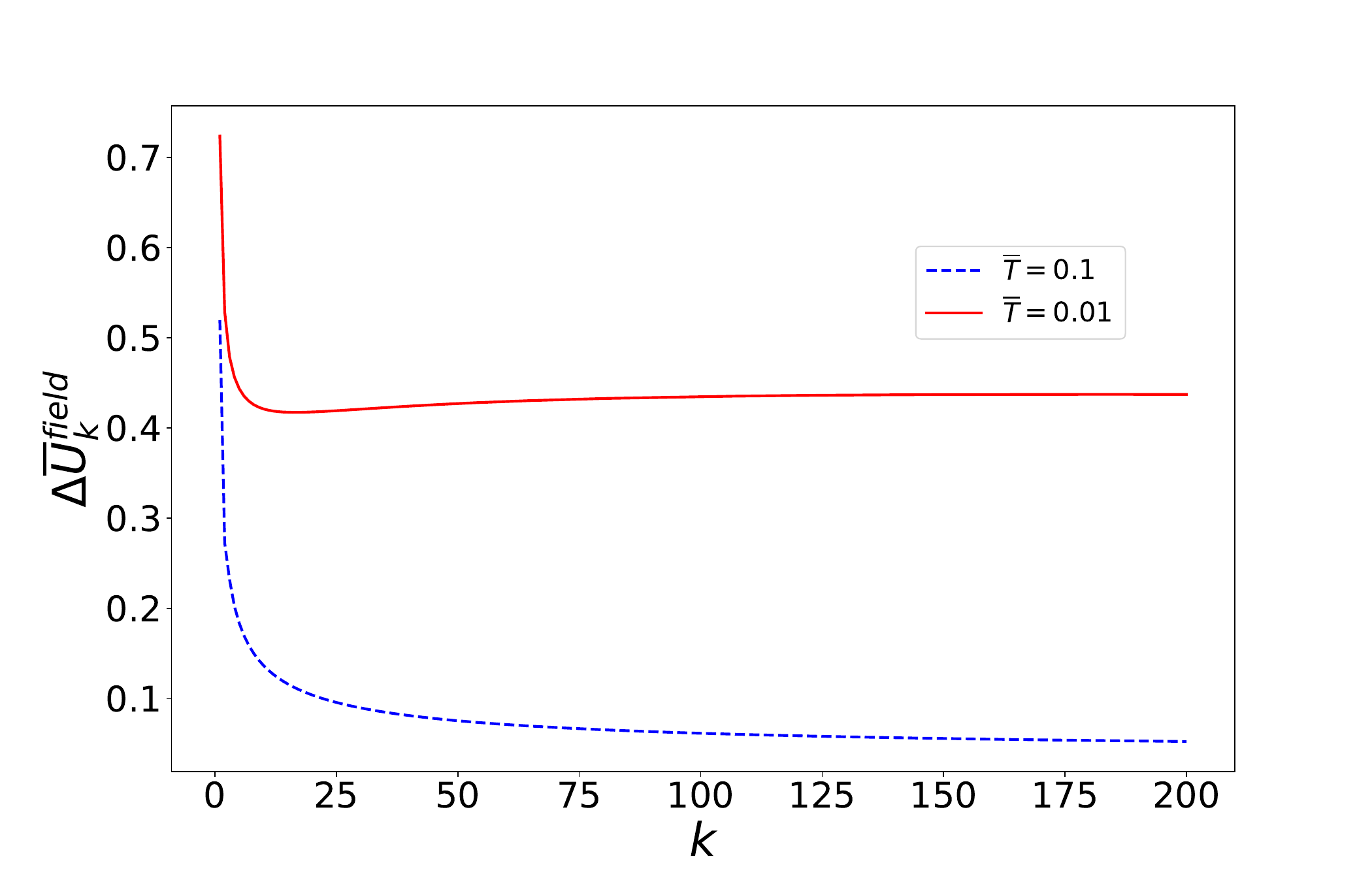}
   \caption{Plot of the $E.M.$ energy gained per collision for two different temperatures. Here, $\bar{U} = U/\hbar\omega_m$  and $\bar{T} = k_BT/\hbar \omega_m$ are the dimensionless energy variation and dimensionless temperature. Parameters: $\omega_{q} = 0.99  \omega_m$, $\Omega_L/g_q = 30$, $\Delta/2\pi = 10^6 Hz$, $g_q = \Delta/600$ and $\omega_m/2\pi = 10^{12} Hz$.} \label{dUf}
\end{figure}

\begin{figure}
  \begin{subfigure}{0.45\textwidth}
    \includegraphics[width=\linewidth]{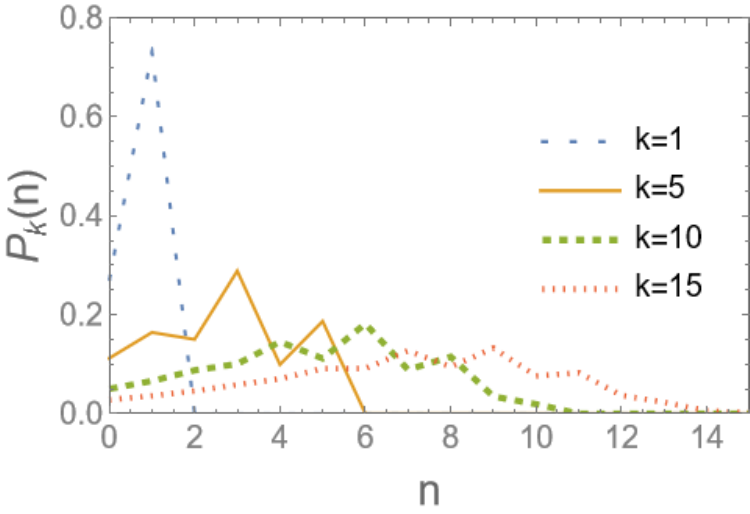}
 \label{PopFC:a}
  \end{subfigure}
  \vspace{\fill}
  \begin{subfigure}{0.45\textwidth}
    \includegraphics[width=\linewidth]{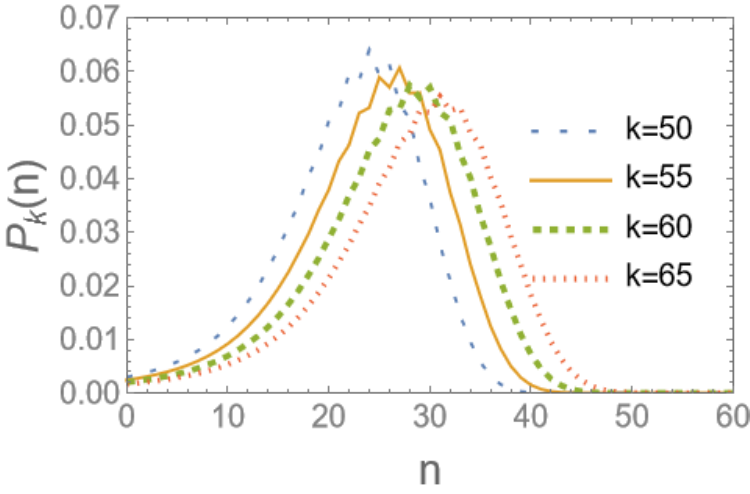}
 \label{PopFC:b}
  \end{subfigure}
  \vspace{\fill}
  \begin{subfigure}{0.45\textwidth}
    \includegraphics[width=\linewidth]{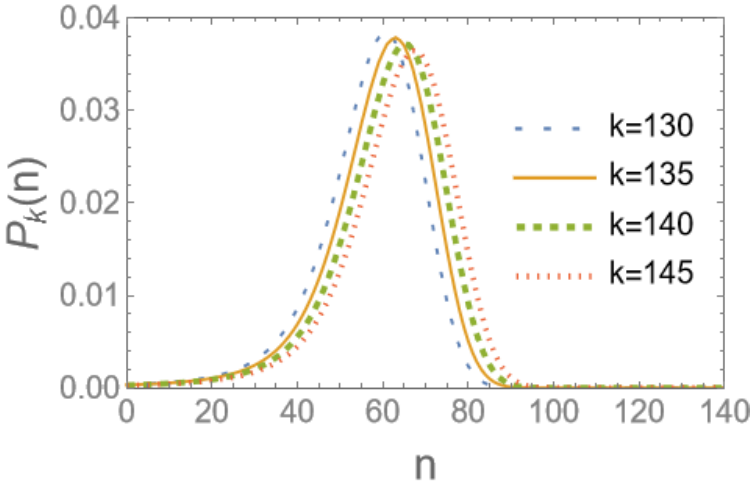}
 \label{PopFC:c}
  \end{subfigure}

\caption{Three plots performed at the same dimensionless temperature $\overline{T}=0.01$ for the case of uncorrelated atomic chain. The number of collisions in each plot is specified in the legend.} \label{PopFC}
\end{figure}

In Fig. $\ref{dUf}$, we plot the field's energy gain per collision which, from Eqs. \eqref{uat} and \eqref{uf}, has the same behavior as its atomic counterpart. Notice that the protocol works better for low temperatures, i.e., the energy gain increases when the temperature decreases. This is explained by the fact that in each interaction there is a population inversion between the energy levels in each doublet. The transfer of population in each collision is more effective, and hence, the population inversion larger, the lower the temperature of the systems, generating, thus, more energy gain. We also see that the energy gain per collision converges to a finite asymptotic value greater than zero. This can be understood by verifying that the charging process for the stationary battery (the E.M. field in our example) slowly drags it into an asymptotic distribution. The high degree of disorder of the initial state of the uncorrelated atomic chain is responsible for raising the populations of different levels of the E.M. field's state at each collision. More than that, the fact that every collision is similar and independent ends up generating a Poisson distribution in the E.M. field as can be checked in Fig. \ref{PopFC}. After a certain number of collisions, which varies with the temperature of the initial state, the E.M. field converges into a large coherent state. This state, for the purpose of the atomic dynamics, works as a classical field, meaning that, after enough collisions with uncorrelated atoms, all the following collisions just implement Rabi flips in the atoms and slightly displace the E.M. field.

\begin{figure}[h!]
    \centering
    \includegraphics[scale=0.7]{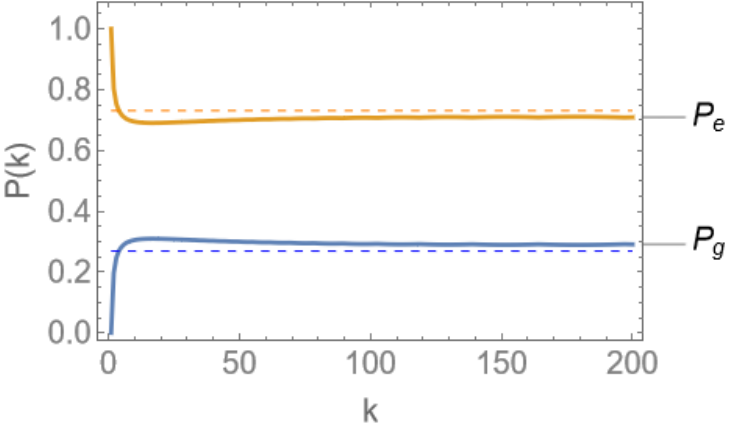}
    \caption{The dashed lines represent the Rabi population inversion when an atom interacts with a classical field. The asymptotic behaviour of the field into a coherent state induces an also asymptotic behavior in the final state's populations of each atom after interaction with the FC.}
    \label{WayToClassical}
\end{figure}

This semi-classical behavior is specifically due to the uncorrelated nature of the atomic chain's global state and is not present when correlations are introduced in the system as we see in the following session.

\section{Correlation in the initial state of the atomic chain}

So far, we have discussed the charging process for an uncorrelated atomic chain. From now on, we will analyze the impact on the charging protocol of considering correlated atoms in the assembling line in two possible scenarios: that of a classically correlated ($C.C$) or entangled ($EN$) atomic chain. In both cases, each atom of the assembling line will still be in a Gibbs state at the same finite temperature $T$ given by \eqref{gs} which allows us to immediately compare the new results with those obtained for the uncorrelated chain.

\subsection{Classically correlated and entangled state}
\label{A}
We take, for the classically correlated streamline of atoms, the density operator
\begin{equation}\label{ccs}
    \rho_{atom}^{C.C}(0) = p_{g}^{T}\ket{gg...}\bra{gg...} + p_{e}^{T}\ket{ee...}\bra{ee...},
\end{equation}
whereas, for the entangled streamline, the GHZ-like global state
\begin{equation}
    \ket{\psi_{atom}(0)} = \sqrt{p_{g}^{T}}\ket{ggg...} + \sqrt{p_{e}^{T}}\ket{eee...}.
    \label{GHZ}
\end{equation}
\noindent
In both cases, the coefficients $\{p_g^T,p_e^T\}$ are chosen such that each individual atom is in the same thermal state as before. The initial state of the hybrid battery (atomic chain + field) will be given by $\rho(0) = \rho_{atom}^{j}(0) \otimes \rho_{field}(0)$, where $\rho_{field}(0)$ is given by Eq. \eqref{gs} and $\rho_{atom}^{j}$ is either $\rho_{atom}^{C.C.}$ or $|\psi_{atom}(0)\rangle\langle \psi_{atom}(0)|$.

\begin{figure}[h!]
  \centering
  \includegraphics[width=1\columnwidth]{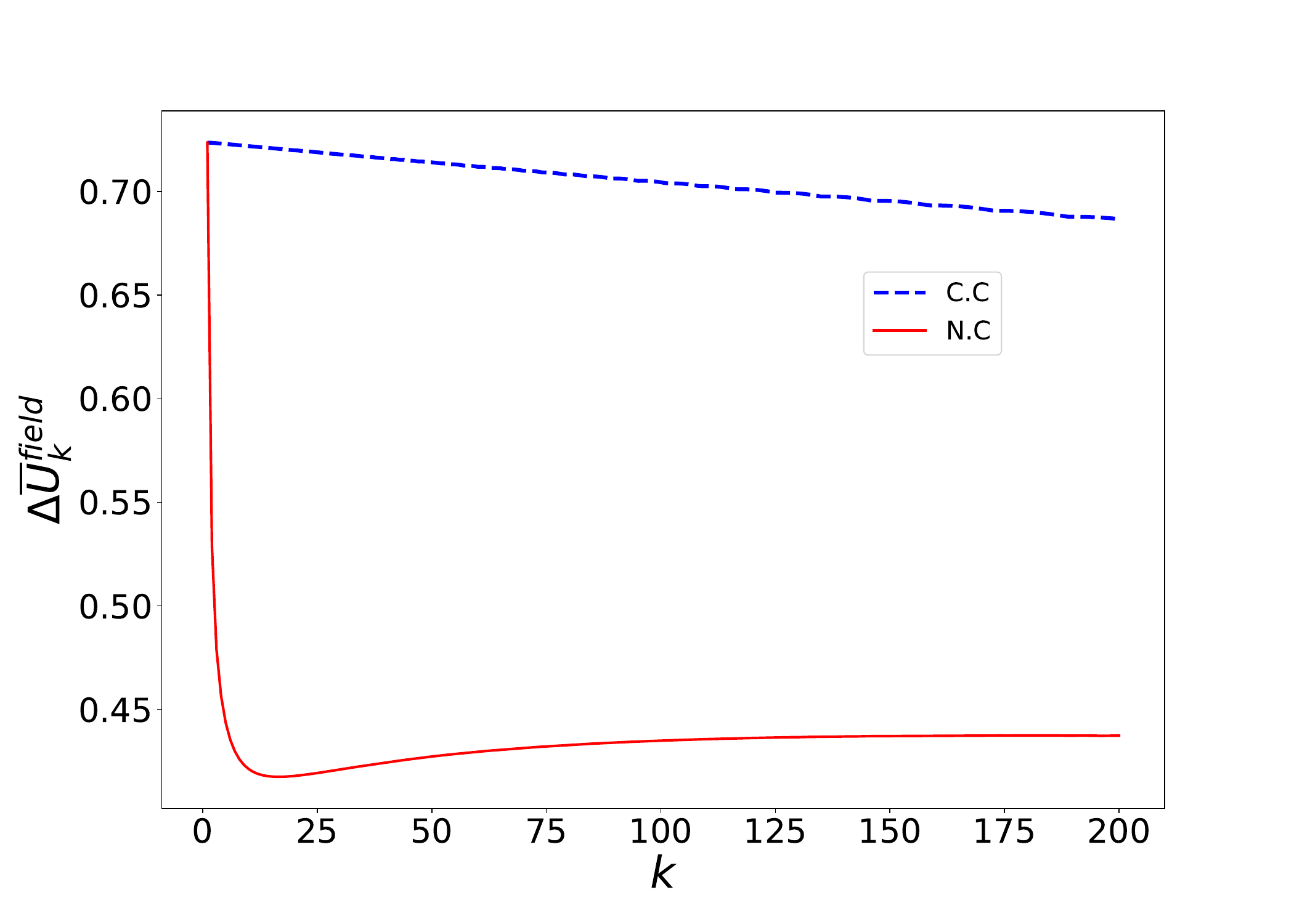}
\caption{Here we plot the energy gain per collision for the E.M. considering two different initial states for the atoms in the assembling line: a classically correlated state ($C.C$) and an uncorrelated state ($N.C$). Parameters: $\omega_{q} = 0.99  \omega_m$, $\Omega_L/g_q = 30$, $\Delta/2\pi = 10^6 Hz$, $g_q = \Delta/600$, $\omega_m/2\pi = 10^{12} Hz$ and $\bar{T} = 0.01$.}
\label{energyCC}
\end{figure}

To show how the presence of correlation in the initial state of the atomic system affects the performance of the charging protocol, we plot in Fig.~\ref{energyCC} the field's energy gain per collision in both cases when the atoms are uncorrelated, and when the atoms are initially in the C.C state \eqref{ccs}. As in the previous section, the energy gained in each stroke by the colliding atom and the field is still determined by the same expression (Eq.\ref{Sk}) and the overall split of the gained energy between the atom and the E.M. field still depends on the the ratio $\omega_{eg}/\omega_q$. 

It is clear that the correlation in the atomic streamline increases the energy gain of the E.M. field. This can be understood by the fact that the correlation reinforces the population transfer to higher number states: atoms in the ground state are the ones that allow for the largest overall gain in the anti-J.C. scenario, and the probability of having N sequential atoms in level $|g\rangle$ is $p_g^T$ for the correlated case and $(p_g^T)^N$ for the uncorrelated one. In fact, the number of ground state atoms for the uncorrelated case follows a binomial distribution of rate $p_g^T$ and the only way for both overall probabilities to coincide is at $T=0$ ($p_g^T=1$).

The same result holds for the entangled atomic chain. This is explained by considering a simple example of only two atoms in the chain ($K=2$) and assuming that the temperature is small enough for the population of the E.M. field to be concentrated in the ground state $\ket{0}$, i.e., $k_BT \ll \hbar \omega_q$. In this scenario, we can easily neglect the field's populations $p_{n}$ for $n\geq 1$ (without any further impact on the atomic chain's state), so that the total system's initial state is given by
\begin{equation}
    \ket{\psi(0)} = \left(\sqrt{p_{g}^{T}}\ket{gg} + \sqrt{p_{e}^{T}}\ket{ee}\right)\otimes\ket{0}.
    \label{fieldEN}
\end{equation}
In the first collision, the state $\ket{g0}$ undergoes a Rabi oscillation and, since the protocol is set to maximize the energetic gain, we calibrate the interaction to stop whenever $\ket{g0} \rightarrow \ket{e1}$. The first collision takes $|\psi(0)\rangle$ into state $|\psi(1)\rangle = \ket{e}\left(\sqrt{p_{g}^{T}}|g\rangle|1\rangle + \sqrt{p_{e}^{T}}\ket{e}\ket{0}\right)$ and the second collision takes $|\psi(1)\rangle$ into $|\psi(2)\rangle = \ket{ee}\left(\sqrt{p_{g}^{T}}|2\rangle + \sqrt{p_{e}^{T}}\ket{0}\right)$. The second atom starts at the partial state $\ket{g1}$ and must undergo a Rabi-flip to $\ket{e2}$. Notice how, at each stroke, the time window of the protocol is dictated solely by the $\ket{gg}$ part of the initial state, since the presence of the atomic excited state $\ket{ee}$ shields the population of level $\ket{0}$ (the state $\ket{e0}$ is an eigenstate of the effective Hamiltonian (\ref{Heff})). All the atoms end up in the excited state, maximizing their charging and the E.M. field ends up on a superposition of the vacuum, and a number state equal to the number of collisions.

A similar calculation carried on for the $C.C.$ case generates the E.M. field state 
\begin{equation}
    \rho_{field}(\tau_2) = p_{g}^{T}\ket{2}\bra{2} + p_{e}^{T}\ket{0}\bra{0},
    \label{fieldCC}
\end{equation}
while also fully charging the atomic chain. Since the energy gained by the E.M. field only depends on the population of each eigenstate of its free Hamiltonian, (i.e. the coherences that appear in the entangled case do not affect it) both entangled and classically correlated cases are energetically equivalent. That is not to say that both cases are absolutely equivalent as we see in the following section.

\subsection{Work extraction and ergotropy}
Another typical figure of merit to evaluate the performance of charging a quantum battery is the ergotropy~\cite{Erg} stored in the final state, which accounts for the maximum amount of work that can be extracted from the system by a unitary process. It is widely known that correlations can be used to boost the performance of a quantum battery~\cite{Gyhm2022, cor, cor2}, and we have already checked how they impact the energetic transfer in our system. Now, we proceed to analyse the stored ergotropy.

Ergotropy can be thought of, as literature shows, as the difference between quantum and classical relative entropies with respect to the Gibbs state \cite{ergentDiff}. Relative entropy is a measure of how distant two states are from each other, and therefore the ergotropy stored in the system must grow as its density matrix gets distinct from the initial thermal state. Taking into account that the usage of correlated global states speeds up the process of shifting the electromagnetic field towards higher occupation numbers if compared to the uncorrelated case, one should expect that in the former scenario, the ergotropy gained by the field in each collision should outperform its uncorrelated counterpart.

Following reference \cite{ergentDiff}, one defines ergotropy as
\begin{equation}
    \beta\mathcal{E} = S\left(\rho ||\rho^{T}\right) - D\left(\rho ||\rho^{T}\right),
    \label{ergotropyAsDif}
\end{equation}
\noindent where $S\left(\rho ||\rho^{T}\right) \equiv Tr\{\rho \left(ln(\rho) - ln(\rho^{T})\right)\}$ and $D\left(\rho ||\rho^{T}\right) \equiv \sum_{i} p_{i} ln\left(\frac{p_{i}}{p_{i}^{T}}\right)$. Here, $\rho$ is the density matrix whose ergotropy we want to obtain, $p_{i}$ are its eigenvalues in descending order (\textit{i.e.}, $p_{i} \geq p_{i+1}$), and $\rho^{T} = exp(-\beta\hat{H})/Z$ is the Gibbs state, with eigenvalues $p_{i}^{T}$. The functions $S\left(\rho ||\rho^{T}\right)$ and $D\left(\rho ||\rho^{T}\right)$ are, respectively, the quantum and classical relative entropies between the desired quantum state and the thermal state.

A simple calculation with both quantities leads to
\begin{eqnarray}
\begin{split}
&S\left(\rho ||\rho^{T}\right) = ln(Z) + \sum_{i} p_{i} ln(p_{i}) + \beta\sum_{i}p_{i}\bra{p_{i}}\hat{H}\ket{p_{i}},\\
&D\left(\rho ||\rho^{T}\right) = ln(Z) + \sum_{i} p_{i} ln(p_{i}) + \beta\sum_{i} p_{i}E_{i}.
\end{split}
\end{eqnarray}

When applied to the system of a quantum field with $\hat{H} = \hbar\omega_{q}\hat{N}$, $E_{i} = \hbar\omega_{q} i$, equation (\ref{ergotropyAsDif}) results in
\begin{equation}
    \mathcal{E} = \hbar\omega_{q}\sum_{i=0}^{\infty} p_{i} \left(\bra{p_{i}}\hat{N}\ket{p_{i}} - i\right).
    \label{FinalErgotropy}
\end{equation}

The system's ergotropy written in the form of (\ref{FinalErgotropy}) eases the visualization that the positive contribution to the maximal stored work in the field is due to states with higher mean occupation numbers, and increases as they get more and more populated. Therefore the shifting of the field's population towards higher mean occupation numbers is directly linked to the stored amount of extractable work: the faster it shifts, the more ergotropy it stores with a given chain size. It is then desirable that the global state of the atomic chain is correlated, since it contributes to a faster shift of the field's population, and therefore to the increment of the total ergotropy.

A simple algebraic calculation can be performed in the regime in which the population of the E.M. field is concentrated in its ground state, i.e.,  when ($k_BT \ll \hbar \omega_q$). Following the same simple evolution we performed on subsection A, one can easily find that for a chain with $K$ atoms, the final states of the field for the entangled and the classically correlated schemes are, respectively
\begin{eqnarray}
\begin{split}
\rho_{field}^{EN}(\tau_K) =& p_{g}^{T}\ket{K}\bra{K} + p_{e}^{T}\ket{0}\bra{0} +\\
&\sqrt{p_{g}^{T}p_{e}^{T}}\left(\ket{K}\bra{0} + \ket{0}\bra{K}\right),\\
\rho_{field}^{C.C.}(\tau_K) = &p_{g}^{T}\ket{K}\bra{K} + p_{e}^{T}\ket{0}\bra{0}.
\end{split}
\end{eqnarray}

Diagonalizing the first state to find its eigenvectors and using formula (\ref{FinalErgotropy}), one finds that the battery's ergotropies for each case are given by
\begin{eqnarray}
\begin{split}
&\mathcal{E}_{EN} = \dfrac{K\hbar\omega_{q}}{1+\frac{p_{e}^{T}}{p_{g}^{T}}},\\
&\mathcal{E}_{CC} = \hbar\omega_{q}(Kp_{g}^{T} - p_{e}^{T}),
\end{split}
\end{eqnarray}
\noindent and its ratio is given by
\begin{equation}
    \dfrac{\mathcal{E}_{EN}}{\mathcal{E}_{CC}} = \dfrac{K}{K - \frac{p_{e}^{T}}{p_{g}^{T}}} = \dfrac{K}{K - e^{-\beta\hbar\omega_{eg}}},
    \label{RatioErgs}
\end{equation}
\noindent which approaches unit as $T\rightarrow 0$, for whatever chain size used. Furthermore, even at non-zero, but still low, temperatures, the ergotropic gain ratio diminishes with longer atomic streams. Altogether, these results illustrate how the entangled state has effectively no considerable gain upon a classical correlation between the atoms.

\begin{figure}[h!]
\centering
\begin{subfigure}{0.55\textwidth}
  \centering
  \includegraphics[width=0.9\linewidth]{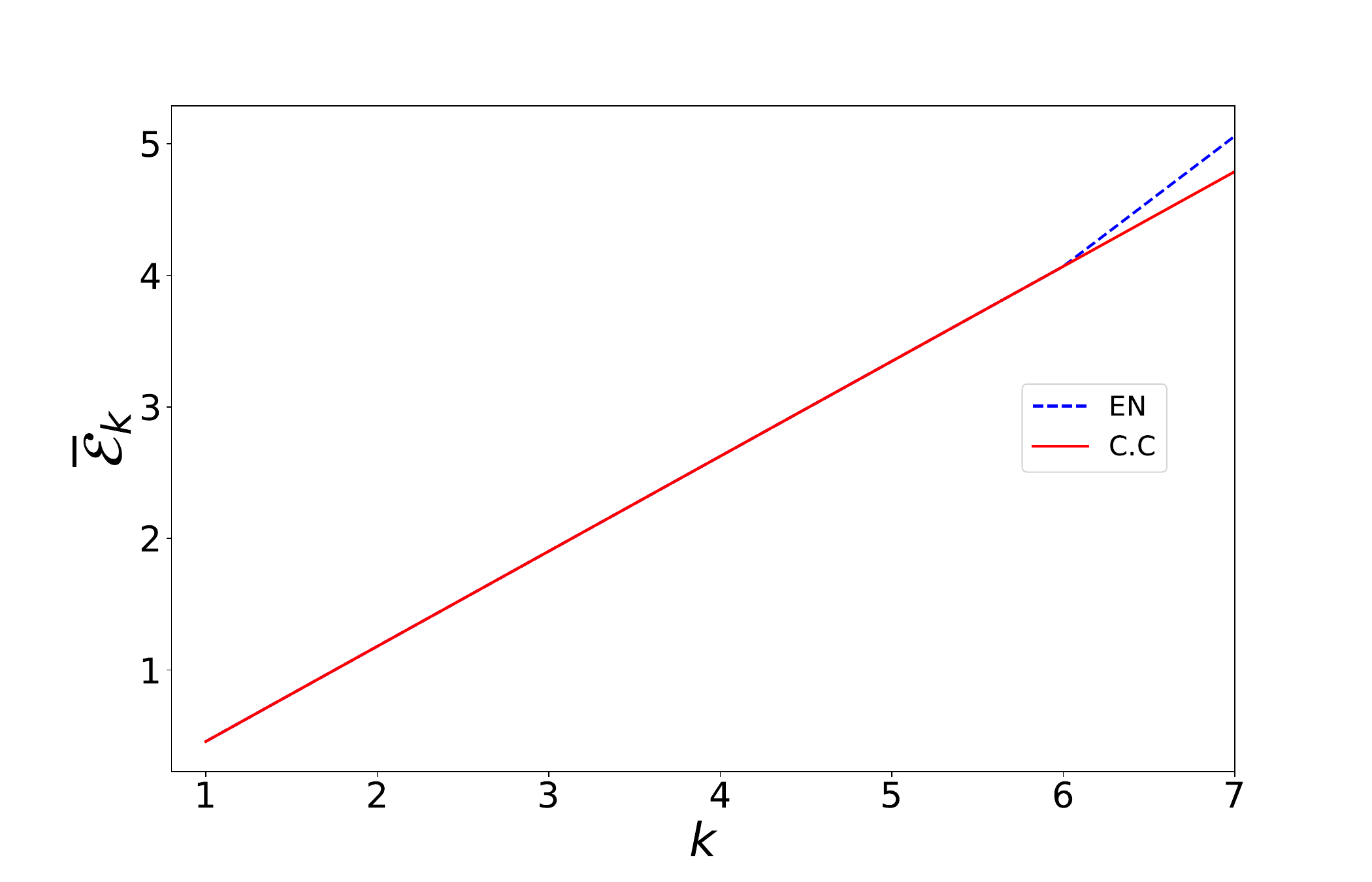}
\end{subfigure}
\begin{subfigure}{0.55\textwidth}
  \centering
  \includegraphics[width=0.9\linewidth]{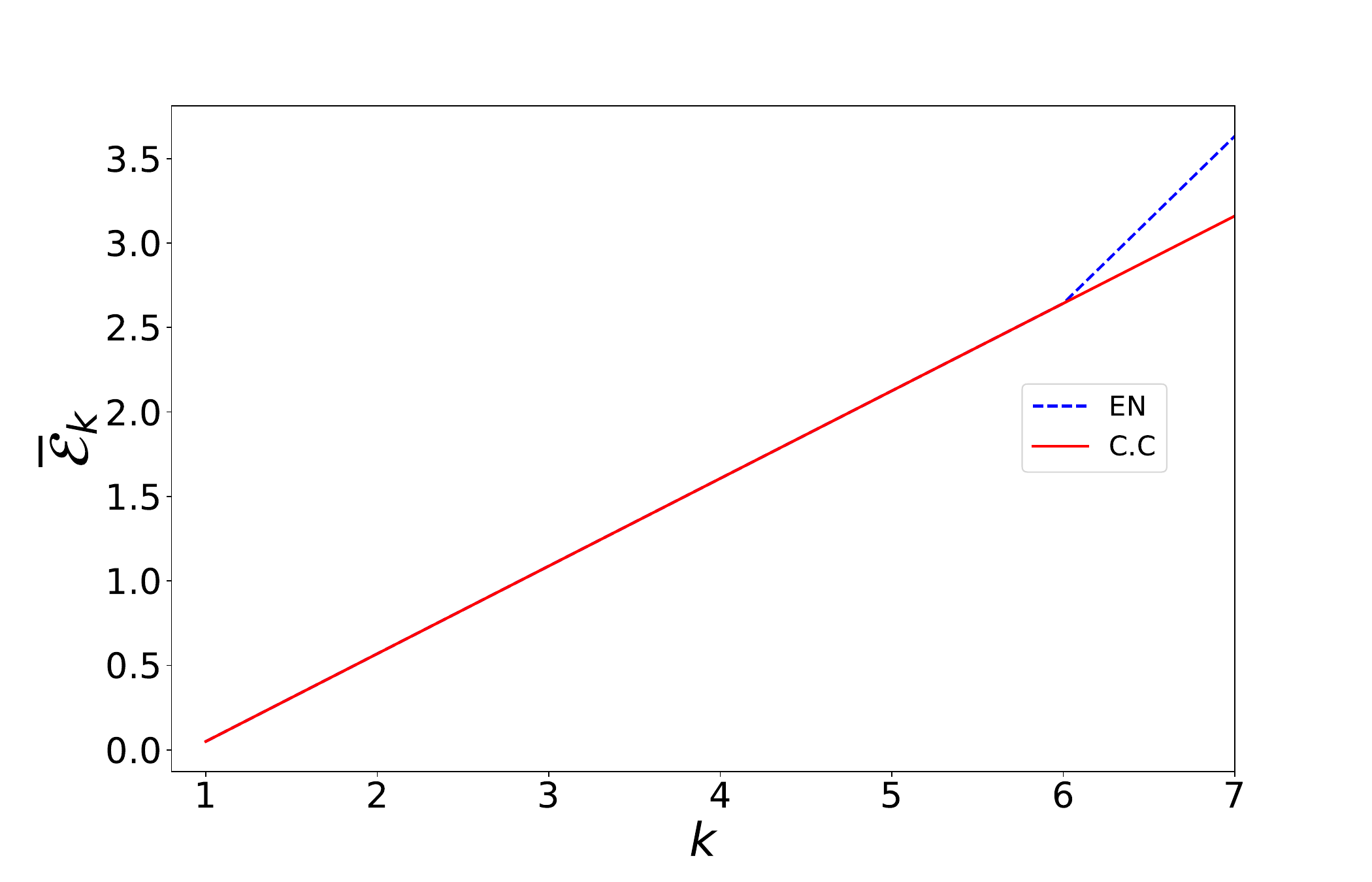}
\end{subfigure}
\caption{Dimensionless ergotropies (normalized by $\hbar\omega_m$) in the field for the entangled and classically correlated initial atomic states, for $\overline{T}=0.01$ (upper) and $\overline{T}=0.1$ (lower). Despite the fact that the entangled initial state increases the amount of ergotropy stored in the field, this increment does not substantially scale with temperature and may even present losses when the cost for producing such a state is considered. Furthermore, notice that this increment is only available after the final atom collides with the field, which is expected since only then the information about the entanglement of the whole chain state is furnished.}
\label{ergotropyCCEN}
\end{figure}

For arbitrary values of the temperature, though, evaluating the ergotropic difference between the two types of correlation is a more involved task. This is so because in this case we can no longer neglect the non-vacuum populations of the field, which implies that more than a single doublet should evolve during each collision, and the optimal time for maximizing the energy extraction must be found numerically. Furthermore, solving the evolution equations for the entangled initial state is a hard computational task, since the total Hilbert space dimension scales exponentially with the number of inputs. In this paper we compare both cases with a chain size of up to seven atoms. Even at this regime, the main source of additional ergotropy is the coherence present in the entangled state but, as we display in Fig. \ref{ergotropyCCEN}, this increment poses no significant gain over the C.C scenario and from the point of view of resource theory may not even cover the total entropic cost for creating the initial entangled state.

Since the entangled global state is a pure state, the entropic cost for creating such a resource is maximal among the ones we used in the protocol. For the classically correlated atomic state, on the other hand, the Von Neumann entropy can be shown to be $S(\rho_{CC}) = \dfrac{\chi}{\overline{T}}\dfrac{e^{-\chi/\overline{T}}}{Z} + ln(Z)$, with $\omega_{g} = 0, \omega_{e} = \chi\omega_{m}$ and $\overline{T} \equiv \frac{k_{B}T}{\hbar\omega_{m}}$ the dimensionless temperature. In order to maintain the population of level $\ket{m}$ under control, to avoid spoiling the adiabatic elimination condition, we should refrain from increasing $\overline{T}$ above a few decimal points. Indeed, in the present case we have limited it to the maximal value of $\overline{T} = 0.1$, which nears the $p_{m}^{T}$ population of each atom to be five orders of magnitude lower than $p_{g}^{T}$. We consider this to be a safe limit for the operation of the protocol, which also grants us the additional perk of barely neglecting non-vacuum field's populations. This is only true, however, if we work under the assumption of low atomic gap $\chi$.

As one may see from the expression for $S(\rho_{CC})$, the reduced costs with resource are achieved for low values of $\chi/\overline{T}$, and therefore raising the temperature (within the protocol's limits) makes the C.C. initial state a better resource, since we reduce the cost for production and, from equation (\ref{RatioErgs}), the entangled state presents no absolute gain for longer chains. The single scenario where the entangled state may outperform the classically correlated case is for small-sized atomic chains, where the entropic gain for the field in the former setup can be up to twice the amount of work the field stores in the latter.

\section{Conclusion}

In this paper we have analysed a protocol where a quantum single-mode of the electromagnetic field is coupled to a stream of atoms with both systems being pumped by an external classical power source in an anti-Jaynes-Cummings configuration. We have considered both the quantum field and the atomic chain as quantum batteries that are simultaneously charged by the external classical drive and have focused on studying the role played by correlations in the atomic chain in the performance of the charging protocol.

In particular, we have shown that while correlations in the atomic system do not impact the charging dynamics of each individual atom, they do improve the charging of the quantized field by decreasing the total time required to reach a certain level of internal energy. This corresponds to an overall increase in the power efficiency of the charging process when compared to the uncorrelated case. Furthermore, the total amount of stored energy and ergotropy in the field are also higher when correlation is used.

We have also compared the effects of purely quantum (entanglement) versus purely classical correlations. In particular, we have tested the collection of atoms both in a GHZ-like entangled state, with appropriate coefficients to mimic individual atoms in a thermal state, or in its classically correlated equivalent (the equivalent mixture of all atoms up or all atoms down). We have shown that, in our protocol, even though entanglement may generate slightly higher values for the ergotropy of the quantum field, this extra gain certainly does not compensate for the much larger entropic cost to produce GHZ-like states versus their much ``cheaper'' classically correlated versions. Furthermore, we have also shown that this small extra gain introduced by entanglement becomes less relevant the larger the chain is, vanishing for infinite ones.

Finally, it is worth noticing that further developments of these results may also consider atom-atom interactions within the chain, in the spirit of the work done in~\cite{interactions}. Such interactions could be used to change the correlations between different pairs of atoms, altogether with their interaction with the mode of the electromagnetic field. In this case, an interesting program would be to analyse how figures of merit such as total charging or the power efficiency of charging the quantum field could be affected by such interactions.

\acknowledgements
This work was supported by  CNPq Projects 302872/2019-1, INCT-IQ 465469/2014-0, and FAPERJ project E-26/202.576/2019. TFFS and YVA thank Capes for their financial support.


\begin{thebibliography}{99}
\bibitem{karen2013}
K. V. Hovhannisyan, M. Perarnau-Llobet, M.s Huber, and A. Acín, Entanglement Generation is Not Necessary for Optimal Work Extraction, Phys. Rev. Lett. {\bf 111}, 240401 (2013).

\bibitem{andolina2018}
G. M. Andolina, D. Farina, A.a Mari, V. Pellegrini, V. Giovannetti, and M.o Polini, Charger-mediated energy transfer in exactly solvable models for quantum batteries, Phys. Rev. B {\bf 98}, 205423 (2018).

\bibitem{Farina2019} D. Farina, G. M. Andolina, A. Mari, M. Polini, V. Giovannetti, Charger-mediated energy transfer for quantum batteries: An open-system approach, Phys. Rev. B {\bf 99}, 035421 (2019).

\bibitem{andolina2019} G. M. Andolina, M. Keck, A. Mari, M. Campisi, V.Giovannetti, and M. Polini, Extractable Work, the Role of Correlations, and Asymptotic Freedom in Quantum Batteries, Phys. Rev. Lett. {\bf 122}, 047702 (2019).

\bibitem{carrega2020} M. Carrega, A. Crescente, D. Ferraro and M. Sassetti, Dissipative dynamics of an open quantum battery, New J. Phys. {\bf 22}, 083085 (2020).

\bibitem{caravelli2020}
F. Caravelli, G. Coulter-De Wit, L. P. García-Pintos, and A. Hamma, Random quantum batteries, Phys. Rev. Research {\bf 2}, 023095 (2020).

\bibitem{sergi2020}
S. Juli\`{a}-Farré, T. Salamon, A. Riera, M. N. Bera, and M. Lewenstein, Bounds on the capacity and power of quantum batteries, Phys. Rev. Research {\bf2}, 023113 (2020).

\bibitem{liu2021}
JX. Liu, HL. Shi, YH. Shi, XH Wang, and WL. Yang, Entanglement and work extraction in the central-spin quantum battery, Phys. Rev. B {\bf 104}, 245418 (2021).

\bibitem{stella2021} S. Seah, M. Perarnau-Llobet, G. Haack, N. Brunner, and S. Nimmrichter, Quantum Speed-Up in Collisional Battery Charging, Phys. Rev. Lett. {\bf 127}, 100601 (2021).

\bibitem{Mondal2022}
S. Mondal and S. Bhattacharjee, Periodically driven many-body quantum battery, Phys. Rev. E {\bf 105}, 044125 (2022).

\bibitem{Gyhm2022} JY. Gyhm, D. Šafránek, and D. Rosa, Quantum Charging Advantage Cannot Be Extensive without Global Operations, Phys. Rev. Lett. {\bf 128}, 140501 (2022).

\bibitem{Ghosh2022}
S. Ghosh and A. Sen, Dimensional enhancements in a quantum battery with imperfections, Phys. Rev. A {\bf 105}, 022628(2022).

\bibitem{Kang2022} C.-K. Hu, J. Qiu, P.J.P. Souza et al., Optimal charging of a
superconducting quantum battery, Quantum Sci. Technol. {\bf7} (4), 045018 (2022).

\bibitem{raf2023}R. Salvia, M. Perarnau-Llobet, G. Haack, N. Brunner, and S. Nimmrichter, Quantum advantage in charging cavity and spin batteries by repeated interactions, Phys. Rev. Research {\bf 5}, 013155 (2023).

\bibitem{dou2023} F.-Q. Dou and F.-M. Yang, Superconducting transmon qubit-resonator quantum battery, Phys. Rev. A {\bf 107}, 023725 (2023).

\bibitem{Horne2020} N. Van Horne, D. Yum, T. Dutta, P. H\"{a}nggi, J. Gong, D. Poletti, and M. Mukherjee, Single-atom energy-conversion device with a quantum load, npj Quantum Information {\bf6}, 37 (2020).

\bibitem{Quach2022} James Q. Quach, Kirsty E. McGhee, Lucia Ganzer, Dominic M. Rouse, Brendon W. Lovett, Erik M. Gauger, Jonathan Keeling, Giulio Cerullo, David G. Lidzey, and Tersilla Virgili. Superabsorption in an organic microcavity: Toward a quantum battery. Science advances, {\bf8(2)}:eabk3160, (2022).

\bibitem{Chang2022} Chang-Kang Hu et al, Optimal charging of a superconducting quantum battery, Quantum Sci. Technol. {\bf7}, 045018 (2022).

\bibitem{joshi2022} J. Joshi and T. S. Mahesh, Experimental investigation of a quantum battery using star-topology NMR spin systems, Phys. Rev. A, {\bf 106}, 042601 (2022).

\bibitem{ScovilDuBois}  H. E. D. Scovil and E. O. Schulz-DuBois, Three-Level Masers as Heat Engines, Phys. Rev. Lett. {\bf 2}, {262} (1959).

\bibitem{60years} W. Niedenzu, M. Huber, and E. Boukobza, Concepts of work in autonomous quantum heat engines, Quantum {\bf 3}, 195 (2019).

\bibitem{60years2} J. Goold, M. Huber, A. Riera, L. D. Rio, and P. Skrzypczyk,  The role of quantum information in thermodynamics—a topical review. J. Phys. A: Math. and Theor. {\bf 49}, 143001 (2016).

\bibitem{60years3} H. T. Quan, YX. Liu, C. P. Sun, and F. Nori, Quantum thermodynamic cycles and quantum heat engines, Phys. Rev. E {\bf 76}, 031105 (2007).

\bibitem{60years4} R. Kosloff, Quantum Thermodynamics: A Dynamical Viewpoint, Entropy 15({\bf 6}), 2100 - 2128 (2013).

\bibitem{60years5} J. P. Pekola, Towards quantum thermodynamics in electronic circuits, Nature Phys. {\bf 11}, 118–123 (2015).

\bibitem{SRQHE} A. Ü. C. Hardal, Ö. E. Müstecaplıoğlu, Superradiant quantum heat engine, Sci. Rep. {\bf 5}, 12953 (2015).

\bibitem{AC} A. C. Santos, B. Çakmak, S. Campbell, N. T. Zinner, Stable adiabatic quantum batteries, Phys. Rev. E {\bf 100}, 032107 (2019).

\bibitem{AC2} M. H. M. Passos, A. C. Santos, M. S. Sarandy, and J. A. O. Huguenin, Optical simulation of a quantum thermal machine, Phys. Rev. A {\bf 100}, 022113 (2019).

\bibitem{molmer} F. Pirmoradian, K. Mølmer, Aging of a quantum battery, Phys. Rev. A {\bf 100}, 043833 (2019).

\bibitem{QD} M. Qutubuddin, K. E. Dorfman,  Incoherent control of optical signals: Quantum-heat-engine approach, Phys. Rev. Research {\bf 3}, 023029 (2021). 

\bibitem{Bera} M. L Bera, S. Julià-Farré, M. Lewenstein, M. N. Bera, Quantum heat engines with Carnot efficiency at maximum power, Phys. Rev. Research {\bf 4}, 013157 (2022).

\bibitem{gemme} G. Gemme, G. M. Andolina, F. M. D. Pellegrino, M. Sassetti, D. Ferraro, Off-Resonant Dicke Quantum Battery: Charging by Virtual Photons, Batteries {\bf 9}, 197 2023.

\bibitem{mm} V. Shaghaghi, V. Singh, G. Benenti, D. Rosa, Micromasers as quantum batteries, Quantum Sci. Technol. {\bf 7}, 04LT01 (2023).

\bibitem{lossy} V. Shaghaghi, V. Singh, M. Carrega, D. Rosa, G. Benenti,  Lossy Micromaser Battery: Almost Pure States in the Jaynes-Cummings Regime. Entropy {\bf 25}, 430 (2023).

\bibitem{sim} G. L. Zanin, T. Häffner, M. A. A. Talarico, E. I. Duzzioni, P. H. Souto Ribeiro, G. T. Landi, and L. C. Céleri, Experimental Quantum Thermodynamics with Linear Optics, Braz. J. Phys. {\bf49}, 783–798 (2019).

\bibitem{sim2} P. Lotshaw, and M. Kellman, Simulating quantum thermodynamics of a finite system and bath with variable temperature, Phys. Rev. E {\bf 100}, 042105 (2019).

\bibitem{sim3} Q. Wu, L. Mancino, M. Carlesso, M. Ciampini, L. Magrini, N. Kiesel, and M. Paternostro, Nonequilibrium Quantum Thermodynamics of a Particle Trapped in a Controllable Time-Varying Potential, PRX Quantum {\bf 3}, 010322 (2022).

\bibitem{sim4} CK. Hu, A. C. Santos, JM. Cui et al, Quantum thermodynamics in adiabatic open systems and its trapped-ion experimental realization, npj Quantum Inf. {\bf 6}, 73 (2020). 

\bibitem{Daemon} N. Cottet, S. Jezouin, L. Bretheau, P.Campagne-Ibarcq, Q. Ficheux, J. Anders, A. Auffèves, R. Azouit, P. Rouchon, and B. Huard, Observing a quantum Maxwell demon at work, Proc. Natl. Acad. Sci. USA {\bf 114}, 7561–7564 (2017).

\bibitem{Daemon2} J. P. Bergfield, S. M. Story, R. C. Stafford, and C. A. Stafford, Probing Maxwell’s Demon with a Nanoscale Thermometer, ACS Nano 7, 4429 (2013).

\bibitem{Daemon3} Yanik, K., Bhandari, B., Manikandan, S., and Jordan, A., Thermodynamics of quantum measurement and Maxwell`s demon`s arrow of time, Phys. Rev. A {\bf 106}, 04222 (2022).

\bibitem{Daemon4} B. Annby-Andersson, P. Samuelsson, V. Maisi, and P. Potts, Maxwell's demon in a double quantum dot with continuous charge detection, Phys. Rev. B {\bf101}, 165404 (2020).

\bibitem{qcomput} J. Parrondo, J. Horowitz, and T. Sagawa, Thermodynamics of information. Nature Phys. {\bf 11}, 131–139 (2015). 

\bibitem{qcomput2} A. Solfanelli, A. Santini, and M. Campisi, Experimental Verification of Fluctuation Relations with a Quantum Computer, PRX Quantum {\bf 2}, 030353 (2021).

\bibitem{qcomput3} S. Toyabe,T. Sagawa, M. Ueda, E. Muneyuki, and M. Sano, Experimental demonstration of information-to-energy conversion and validation of the generalized Jarzynski equality, Nature Phys {\bf6}(12), 988-992 (2010).

\bibitem{qcomput4} J. V. Koski, V. F. Maisi, T. Sagawa, and J. P. Pekola, Experimental observation of the role of mutual information in the nonequilibrium dynamics of a Maxwell demon, Phys. Rev. Lett. {\bf 113}, 030601 (2014).

\bibitem{qcomput5} Plenio, M. B., and V. Vitelli, The physics of forgetting: Landauer's erasure principle and information theory, Cont. Phys., {\bf42}, 25-60 (2021).

\bibitem{charger} L. Gao, C. Cheng, WB. He, R. Mondaini, XW. Guan, and HQ. Lin, Scaling of energy and power in a large quantum battery-charger model, Phys. Rev. Research {\bf 4}, 043150 (2022).

\bibitem{charger2} M. B. Arjmandi, A. Shokri, E. Faizi, and H. Mohammadi, Performance of quantum batteries with correlated and uncorrelated chargers, Phys. Rev. A {\bf106}, 062609 (2022).

\bibitem{charger3} F. T. Tabesh, F. H. Kamin, and S. Salimi, Environment-mediated charging process of quantum batteries, Phys. Rev. A {\bf102}, 052223 (2020).

\bibitem{charger4} FQ. Dou and FM. Yang, Superconducting transmon qubit-resonator quantum battery, Phys. Rev. A {\bf 107}, 023725 (2023).

\bibitem{correl1} M. P. Llobet, K. V. Hovhannisyan, M. Huber, P. Skrzypczyk, N. Brunner, and A. Acín, Extractable Work from Correlations, Phys. Rev. X {\bf 5}, 041011 (2015).

\bibitem{correl2} A. Santini, A. Solfanelli, S. Gherardini, and M. Collura, Work statistics, quantum signatures, and enhanced work extraction in quadratic fermionic models, Phys. Rev. B {\bf 108}, 104308 (2023).

\bibitem{correl3} M. P. Llobet, E. Bäumer, K. V. Hovhannisyan, M. Huber, and A. Acin, No-Go Theorem for the Characterization of Work Fluctuations in Coherent Quantum Systems, Phys. Rev. Lett. {\bf 118}, 070601 (2017).

\bibitem{Erg} A. E. Allahverdyan, R. Balian, and Th. M. Nieuwen-huizen, Maximal work extraction from finite quantum systems, Europhys. Lett. {\bf67}, 565 (2004).

\bibitem{MKR} M. F. Santos, E. Solano, R.L. de Matos Filho, Conditional large Fock state preparation and field state reconstruction in Cavity QED, Phys. Rev. Lett. {\bf87}, 93601 (2001).

\bibitem{MFS} M. F. Santos, Universal and deterministic manipulation of the quantum state of harmonic oscillators: a route to unitary gates for Fock State qubits. Phys. Rev. Lett. {\bf 95}, 010504 (2005).

\bibitem{TYM} T. F. F. Santos, Y. V. Almeida, M. F. Santos, Vacuum enhanced charging of a quantum battery, Phys. Rev. A, {\bf 107}, 032203 (2023).

\bibitem{MFSFB} Z. Beleño, M.F. Santos, F. Barra, Laser powered dissipative quantum batteries in atom-cavity QED, arXiv:2310.09953

\bibitem{Nicim} N. Zagury, A. Arag\~ao, J. Casanova, and E. Solano, Unitary expansion of the time evolution operator, Phys. Rev. A, {\bf82}, 042110 (2010)

\bibitem{cor} J. Q. Quach and W. J. Munro, Using Dark States to Charge and Stabilize Open Quantum Batteries, Phys. Rev. Applied {\bf14}, 024092 (2020).

\bibitem{cor2} T. P. Le, J. Levinsen, K. Modi, M. M. Parish, and F. A. Pollock, Spin-chain model of a many-body quantum battery, Phys. Rev. A {\bf 97}, 022106 (2018).

\bibitem{ergentDiff} A. Sone and S. Deffner. Quantum and Classical Ergotropy from Relative Entropies. Entropy 2021, 23, 1107. https://doi.org/10.3390/e23091107

\bibitem{interactions} D. Rossini, G. M. Andolina, and M. Polini, Many-body localized quantum batteries, Phys. Rev. B {\bf 100}, 115142 (2019).

\end{thebibliography}
\end{document}